\documentclass[prd,nofootinbib,preprint,showpacs,superscriptaddress,
               showkeys]{revtex4}
\usepackage{graphicx}
\usepackage{amssymb}
\def\nn{\nonumber}
\def\ni{\noindent}
\def\vev#1{\left\langle #1\right\rangle}
\def\ifmath#1{\relax\ifmmode #1\else $#1$\fi}
\def\half{\ifmath{{\textstyle{1 \over 2}}}}
\def\vb#1{\vbox to #1 pt{}}

\newcommand{\AddrAHEP}{%
  AHEP Group, Instituto de F\'{\i}sica Corpuscular --
  C.S.I.C./Universitat de Val{\`e}ncia \\
  Edificio Institutos de Paterna, Apt 22085, E--46071 Valencia, Spain}
\newcommand{\AddrLisb}{%
 Departamento de F\'\i sica and CFTP, Instituto Superior T\'ecnico\\
          Av. Rovisco Pais 1, $\:\:$ 1049-001 Lisboa, Portugal }

\begin{document}

\preprint{IFIC/04--63}

\vspace*{2cm} \title{Charge Breaking Minima in the Broken R--parity
  Minimal Supersymmetric Standard Model}

\author{M.~Hirsch} \email{mahirsch@ific.uv.es}\affiliation{\AddrAHEP}
\author{C.~Hugonie}\email{hugonie@ific.uv.es}\affiliation{\AddrAHEP}
\author{J.~C.~Rom\~ao}\email{jorge.romao@ist.utl.pt}\affiliation{\AddrLisb}
\author{J.~W.~F.~Valle} \email{valle@ific.uv.es}\affiliation{\AddrAHEP} 

\vspace*{3mm}

\begin{abstract}  
  We reconsider the possible presence of charge and colour breaking
  minima in the scalar potential of the minimal supersymmetric
  standard model (MSSM) and its minimal generalization with R--parity
  explicitly broken by bilinear terms (RMSSM).  First we generalize
  some results previously derived for the MSSM case. Next we
  investigate how robust is the MSSM against its RMSSM extension.  We
  examine the constraints on the RMSSM parameter space that follow
  from the required absence of charge breaking minima in the scalar
  potential.  We point out the possibility of generating non--zero
  vacuum expectation values for the charged Higgs field which is not
  present in the MSSM.  However, given the smallness of neutrino
  masses indicated by neutrino oscillation data, we show that the
  RMSSM represents only a slight perturbation of the MSSM and is thus
  as safe (or unsafe) as the MSSM itself from unwanted minima in the
  scalar potential.
\end{abstract}

\vspace*{5mm}
 

\maketitle

\newpage

\section{Introduction}

Softly broken supersymmetric models contain a fairly large number of
scalar fields not present in the standard model. Their existence leads
to a complicated scalar potential, which might contain undesirable
minima which spontaneously break charge and/or color symmetry, a
situation which can not happen within the Standard Model. The
condition that the ``realistic'' minimum is the global minimum of the
theory can be used to obtain restrictions on the parameter space of
supersymmetric models, as already realized more than 20 years ago
\cite{Frere:1983ag,Claudson:1983et,Nilles:1982mp}. This way a
disadvantage of supersymmetry may turn into a virtue by shedding some
light into the unknown supersymmetry breaking mechanism itself.

Due to the enormous complexity of the full scalar potential in the
minimal supersymmetric extension of the standard model (MSSM) early
papers on this
subject~\cite{Frere:1983ag,Claudson:1983et,Nilles:1982mp,Komatsu:1988mt}
have only analyzed particular, but especially dangerous directions in
field--space. Casas et al \cite{casas:1997ze} have presented a more
detailed analysis of this subject. They were able to show that in the
constrained MSSM (CMSSM) with minimal supergravity boundary conditions
strong constraints arise ruling out sizeable parts of the parameter
space~\cite{casas:1997ze}.

Similar studies in R--parity violating versions of the MSSM, however,
have not been published~\footnote{The work of Abel and Savoy
  \cite{Abel:1998ie} contains a discussion on the possibility of
  lifting flat directions by adding explicit trilinear R--parity
  violating terms to the superpotential. However, they discuss the
  impact of bilinear terms only briefly. This is our main emphasis.}.
Our main goal is to present a detailed analysis of the
'unbounded--from--below' (UFB) as well as charge/colour breaking
minima (CCB) in the bilinear R--parity breaking model
(RMSSM)~\cite{Diaz:1997xc}.  This model breaks lepton number and
R--parity explicitly through the simplest bilinear terms.  The
justification for such emphasis is threefold.

First, it represents the simplest possible scheme of R--parity
violation, a mere six parameter extension of the MSSM. It is therefore
interesting to investigate the ``stability'' of the MSSM against such
``innocuous'' perturbation.  For this reason we can also call this
model the generalized MSSM where R--parity breaks in the minimal way.
Second, this model is motivated by the fact that it produces the
paradigm for the idea that supersymmetry is the origin of neutrino
mass~\cite{Hirsch:2004he}, leading to a pattern of neutrino
masses~\cite{hirsch:2000ef} that successfully describes current
neutrino data~\cite{Maltoni:foc}.  Last, but not least, it represents
the only model of R--parity breaking consistent with a spontaneous
violation of R--parity~\cite{Masiero:1990uj,Romao:1992vu}, where it is
the vacuum, not the fundamental theory, that breaks the symmetry.

In this model the atmospheric neutrino mass scale \cite{Fukuda:1998mi}
is generated at the tree--level, through the mixing of the three
neutrinos with the neutralinos~\cite{Ellis:1984gi}, in an effective
`low--scale'' variant of the seesaw mechanism. In contrast, the solar
mass and mixings needed to account for solar neutrino
data~\cite{Ahmad:2002jz,Eguchi:2002dm} are generated
radiatively~\cite{hirsch:2000ef}.

A very important difference between such a supersymmetric approach to
the origin of neutrino mass and seesaw--type schemes, is that here the
dimension--five operator responsible for (Majorana) neutrino masses is
generated at an accessibly low energy scale -- namely the weak scale.
This makes this model potentially testable by experiment.
In fact it has been shown that such a low--scale scheme for neutrino
masses has the advantage of being testable also ``outside'' the realm
of neutrino physics experiments. Although neutrino properties can not
be predicted from first principles, interpreting current neutrino data
in this framework implies unambiguous tests of the theory at
accelerator
experiments~\cite{Mukhopad:1998xj,Porod:2000hv,Hirsch:2002ys,Chun:2002rh,Hirsch:2003fe}
which can potentially be used to falsify the model.

This paper is organized as follows. In the next section we will
briefly recall some basics of the discussion on CCB and UFB bounds in
the MSSM.  This will serve as a basis for section 3, where we will
discuss new features related to the R--parity violating terms. We show
how the bounds from unbounded--from--below directions have to be
modified, once non--zero bilinear R--parity violating (BRpV) terms are
allowed. We point out the novel possibility to generate a non--zero
vacuum expectation value of the charged Higgs field, albeit in regions
of parameter space which are now excluded by neutrino physics
\cite{Maltoni:foc}. We show that, given current data on neutrino
masses, bilinear R--parity violation can be understood as a small
perturbation of the MSSM. From the point of view of charge breaking
minima the RMSSM is thus as safe (or unsafe) as the MSSM itself. We
will then close with a short summary.

\section{Review of the MSSM results on UFB and CCB}

\noindent
To set up the notation, the superpotential of the MSSM can 
be written as
\begin{eqnarray}
W&=&\varepsilon_{ab}\left[ 
 h_U^{ij}\widehat Q_i^a\widehat U_j\widehat H_u^b 
+h_D^{ij}\widehat Q_i^b\widehat D_j\widehat H_d^a 
+h_E^{ij}\widehat L_i^b\widehat R_j\widehat H_d^a 
-\mu\widehat H_d^a\widehat H_u^b
 \right].
\label{eq:W}
\end{eqnarray}
Here, $h_U^{ij}$, $h_D^{ij}$ and $h_E^{ij}$ are $3\times 3$ Yukawa 
matrices, $\widehat Q$, $\widehat U$ and $\widehat D$ are quark doublet and 
singlet superfields and $\widehat L$ and $\widehat R$ are the usual lepton 
doublet and singlet fields. 
Supersymmetry must be broken and the most general set of soft breaking terms 
allowed by the standard model gauge group under the assumption of 
lepton number conservation can be written as
\begin{eqnarray} 
\label{Vsoft}
{V}_{SB}&\hskip-5mm=\hskip-5mm& 
M_Q^{ij2}\widetilde Q^{a*}_i\widetilde Q^a_j+M_U^{ij2} 
\widetilde U_i\widetilde U^*_j+M_D^{ij2}\widetilde D_i 
\widetilde D^*_j
+M_L^{ij2}\widetilde L^{a*}_i\widetilde L^a_j 
+M_R^{ij2}\widetilde R_i\widetilde R^*_j
+\sum_{i=1}^2 m_{H_i}^2 H^{a*}_i H^a_i \cr
&&+\left[
- \half \sum_{i=1}^3 M_i\lambda_i\lambda_i
+\varepsilon_{ab}\left( 
A_U^{ij}h_U^{ij}\widetilde Q_i^a\widetilde U_j H_u^b 
+A_D^{ij}h_D^{ij}\widetilde Q_i^b\widetilde D_j H_d^a 
+A_E^{ij}h_E^{ij}\widetilde L_i^b\widetilde R_j H_d^a 
\right.\right.\cr
&&\left. \left.  \hskip 45mm
-B\mu H_d^a H_u^b 
\right) +h.c. \vb{18} \right]
\end{eqnarray} 
The Higgs doublets giving mass to the standard model fermions are 
\begin{equation}
  \label{eq:2}
  H_d=\left(
    \begin{array}{l}
      H_d^0\\
      H_d^-
    \end{array}
    \right),
    \qquad
  H_u=\left(
    \begin{array}{l}
      H_u^+\\
      H_u^0
    \end{array}
    \right)
\end{equation}
and the parameters in Eq. (\ref{Vsoft}) are to be understood at some 
renormalization scale $Q$ chosen to minimize the effects of the one loop 
corrections. This way we can neglect in the analysis the effect of the one
loop radiative corrections \cite{casas:1997ze}.
Without loss of generality, we now consider that the fields take the
following vev's\footnote{Our normalization here for the vev's differs
  from Refs. \cite{Diaz:1997xc,hirsch:2000ef} by a factor of
  $\sqrt{2}$.},
\begin{equation}
  \label{eq:3}
  \vev{H_u^+}=0,\quad   \vev{H_d^-}=v_{-},\quad   \vev{H_d^0}=v_d,\quad 
  \vev{H_u^0}=v_u
\end{equation}
to obtain 
\begin{eqnarray}
  \label{eq:4}
  V_{\hbox{Higgs}}&=&\left(m^2_{H_u} + \mu^2\right) v_u^2 
+\left(m^2_{H_d} + \mu^2\right) \left(v_d^2 +v_{-}^2 \right)  
- 2 B \mu v_u v_d 
 -\frac{1}{2} g^2 v_u^2 v_d^2\nn \\[+2mm]
&&+
\frac{1}{8} \left(g^2 +g'^2\right) \left(v_u^4+v_d^4+v_{-}^4+2 v_d^2
  v_{-}^2\right) 
+\frac{1}{4} \left(g^2 -g'^2\right) \left( v_d^2 + v_{-}^2\right) v_u^2
\end{eqnarray}
This Higgs potential has the property that $v_{-}=0$. To see this 
we note that the potential can be written in the form,
\begin{equation}
  \label{eq:5}
  V_{\hbox{Higgs}}= C_4 v_{-}^4 + C_2 v_{-}^2 + C_0
\end{equation}
where
\begin{eqnarray}
  \label{eq:6}
  C_4&=&\frac{1}{8} \left(g^2+g'^2\right)\nn \\
  C_2&=&\frac{1}{4} \left(g^2-g'^2\right) v_u^2 +\frac{1}{4}
  \left(g^2+g'^2\right) v_d^2 + \left(m^2_{H_d} + \mu^2\right)\nn \\
  C_0&=&\frac{1}{8} \left(g^2+g'^2\right) \left(v_u^2-v_d^2\right)^2 
+\left(m^2_{H_u} + \mu^2\right) v_u^2
+\left(m^2_{H_d} + \mu^2\right) v_d^2
-2 B \mu v_u v_d
\end{eqnarray}
Now since $g>g'$ we must have $C_2 >0$, unless $m^2_{H_d} + \mu^2 <0$
\footnote{Casas et al.~\cite{casas:1997ze} assume that only $m^2_{H_u}
  + \mu^2$ can be negative. Even though in mSugra at very large
  $\tan\beta$ values $m^2_{H_d} + \mu^2 <0$ can occur in exceptional
  cases, we will follow their assumption.}. Therefore the minimum of
the Higgs potential occurs for vanishing vev of the charged Higgs
boson.

By using the minimization equations,
\begin{eqnarray}
  \label{eq:7}
  0&=&-2 B \mu v_d + 2 \left(m^2_{H_u} + \mu^2\right) v_u - \frac{1}{2}
  \left(g^2+g'^2\right)  \left(v_d^2-v_u^2\right) v_u \nn \\[+2mm]
   0&=&-2 B \mu v_u + 2 \left(m^2_{H_d} + \mu^2\right) v_d + \frac{1}{2}
  \left(g^2+g'^2\right)  \left(v_d^2-v_u^2\right) v_d 
\end{eqnarray}
one can find the value of the Higgs potential at the real minimum,
\begin{eqnarray}
  \label{eq:8}
  V_{MIN}&=&-\frac{1}{8} \left(g^2+g'^2\right) 
             \left(v_u^2-v_d^2\right)^2 
\end{eqnarray}
Eq. (\ref{eq:8}) will be important to compare with the values of other 
(and potentially deeper) minima.

Before starting the discussion of the dangerous directions, a word of
caution should be added, namely, that the condition that the realistic
minimum is the global one might actually be too conservative. In fact,
it is possible that the universe resides in a false vacuum which is
stable because the tunneling time into the global minimum is large
with respect to the age of the universe. In this sense, CCB and UFB
constraints on the supersymmetric parameter space are sufficient but
might not be necessary, see for example
\cite{Kusenko:1996jn,Kusenko:1996xt}.  However, we will not follow
this line of reasoning any further.

\subsection{UFB directions}

The 'unbounded--from--below' (UFB) directions are those where the
quartic D--terms vanish and some coefficient(s) quadratic in the vev's
are negative. Then the potential at the weak scale
seems to be unbounded from below.  However, this is a slight
misnomer, since if one assumes that all soft masses are positive at
the high unification scale, it appears that these dangerous directions
are not really unbounded from below but there exists a true local
minimum at some large scale. It then must be checked that this local
minimum is not deeper than the physical one.  As was shown in Ref.
\cite{casas:1997ze} there are three kinds of such directions. The
first and most obvious one corresponds to the D--flat direction where
$|v_u|=|v_d|$, all other vev's being zero. The potential along this
direction reads,
\begin{equation}
  \label{eq:10}
  V_{UFB-1}=\left(m^2_{H_u} + m^2_{H_d} +2 \mu^2 -2 |B \mu| \right) v_u^2
\end{equation}
and a sufficient condition to avoid developping a deep minimum at large 
values of the field is
\begin{equation}
  \label{eq:11}
  m^2_{H_u} + m^2_{H_d} +2 \mu^2 -2 |B \mu| >0.
\end{equation}
In principle, one should check the depth of the true minimum along the 
dangerous direction when this coefficient is negative. For simplicity, we 
will stick however to the condition given in Eq. (\ref{eq:11}).

The second dangerous direction corresponds to the case where a slepton
$L_i$ takes a vev $v_i$. Then a combination of $v_u, v_d$ and $v_i$
can cancel the D--term and the potential reads,
\begin{equation}
  \label{eq:12}
  V_{UFB-2}=\left( m^2_{H_u} + \mu^2 + m^2_{L_i} - \frac{| B
  \mu|^2}{m^2_{H_d}+\mu^2 -m^2_{L_i}} \right) v_u^2 -\frac{2
  m^4_{L_i}}{g^2+g'^2}
\end{equation}
which constrains the coefficient of the quadratic term as
\begin{equation}
  \label{eq:ufb2}
m^2_{H_u} + \mu^2 + m^2_{L_i} - \frac{| B
  \mu|^2}{m^2_{H_d}+\mu^2 -m^2_{L_i}} > 0.
\end{equation}
Note that in the case of a universal $m_0$ at the unification scale
the $m_{L_i}$ are usually the smallest soft masses at the weak scale.
Dropping the universality assumption the bound obtained for $m_{L_i}$,
Eq. (\ref{eq:ufb2}), must be verified for the squark soft masses as
well.

Finally the last UFB direction corresponds to the case where $v_d=0$
but we have a neutral slepton $L_i$ with nonzero vev, like in the
UFB-2 case. This direction is both D-- and F--flat. The difference
with respect to UFB-2 is that the F--term is canceled by giving vev's
to the charged sleptons. The resulting potential reads
\begin{equation}
  \label{eq:ufb3}
  V_{UFB-3}=\left( m^2_{H_u} + m^2_{L_i} \right) v_u^2 
  + \frac{|\mu|}{h_{e_j}} \left(m^2_{L_i}+ m^2_{L_j}+ m^2_{e_j}\right) v_u 
 -\frac{2 m^4_{L_i}}{g^2+g'^2}
\end{equation}
Since $m^2_{H_u}$ must be negative in order to break electroweak symmetry and
$m^2_{L_i}$ is small when one assumes universality of the soft terms, the
coefficient quadratic in $v_u$ is generally negative. As shown in
Refs.~\cite{casas:1997ze,Abel:1998ie} in the case of universal soft masses at
the GUT  scale, the condition that the minimum along this UFB-3 direction is
not  deeper than the physical minimum implies $m_0 > \alpha M_{1/2}$, where
$\alpha$ is a coefficient of ${\cal O}(1)$.

\subsection{CCB minima}
\label{sec:CCBMSSM}

\noindent 
For the classical CCB minima, dangerous negative contributions to the
scalar potential are generated by cubic ($A$--type) soft supersymmetry
breaking terms. Therefore these directions cannot be F--flat, but they
are still D--flat. The traditional bound of Ref. \cite{Frere:1983ag}
corresponds to the case where
\begin{equation}
  \label{eq:13}
  \vev{Q^1}=\vev{H_u^2}=\vev{U}=v
\end{equation}
all other vev's vanishing. This choice cancels the D--term and the
potential reads,
\begin{equation}
  \label{eq:14}
  V_{CCB}=v^2 \left(3 h_u^2 v^2 + 2 A_u h_u v + m^2_{H_u}+\mu^2+m^2_Q
  +m^2_U \right)
\end{equation}
In order to avoid a very deep color and charge breaking minimum we
must make sure that the parenthesis in Eq.~(\ref{eq:14}) never
vanishes, which happens if the corresponding second order equation can
not have real solutions.  This leads to the well known condition,
\begin{equation}
  \label{eq:16}
  |A_u|^2 < 3 \left( m^2_{H_u}+\mu^2+m^2_Q +m^2_U \right)
\end{equation}
A more complete and general analysis of this and similarly dangerous
directions can be found in Ref. \cite{casas:1997ze}. Note again, that
the bound given in Eq.(\ref{eq:16}) for $A_u$ must be checked for all
$A$--terms in the general non--universal MSSM.


\section{UFB and CCB in the RMSSM}

The RMSSM is simply the bilinear R--parity violating model, defined by
the following superpotential~\cite{Diaz:1997xc}
\begin{eqnarray}
W&=&W_{MSSM} + \varepsilon_{ab}\epsilon_i\widehat L_i^a\widehat H_u^b
\label{eq:Wrpv}
\end{eqnarray}
and corresponding soft supersymmetry breaking terms,
\begin{eqnarray} 
{V}_{SB}&=& V_{MSSM} + 
 B_i\epsilon_i\widetilde L^a_i H_u^b \,.
\label{eq:Vrpv}
\end{eqnarray} 
It is therefore a rather mild extension of the MSSM. In the following
it will be sufficient to consider for simplicity only a one generation
version of the model~\footnote{We do not believe that this
  simplification has any impact on the following discussion, since
  neutrino oscillation data require $\frac{\epsilon}{\mu}\ll 1$ and
  intergenerational effects between different families of leptons due
  to BRpV terms scale as $(\frac{\epsilon}{\mu})^2$.}. We are mainly
interested in studying how the appearance of the new terms in the
superpotential (and in $V_{SB}$) changes the conclusions which hold
for the MSSM. Since the MSSM is the limit of the RMSSM when $\epsilon
\to 0$ we expect that the results of the MSSM will hold in that limit.
Note also that the structure of the trilinear terms is not modified,
so conclusions like those of Eq.~(\ref{eq:16}) are expected also to
hold in our case.  Defining
\begin{equation}
  \label{eq:17}
   \vev{H_u^+}=0,\quad   \vev{H_d^-}=v_{-},\quad   \vev{H_d^0}=v_d,\quad 
  \vev{H_u^0}=v_u, \quad \vev{L^0}=v',  \quad \vev{L^-}=v'_{-}
\end{equation}
one finds for the scalar potential 
\begin{eqnarray}
  \label{eq:15}
  V&=& M^2_{H_u} v_u^2 + M^2_{H_d} \left(v_d^2 + v_{-}^2\right) 
+ M^2_{L} \left( v'^2 +  v'^2_{-}\right) - 2 B \mu\, v_d v_u + 2 B'
\epsilon\, v_u v' \nn\\[+1mm]
&&
+\epsilon^2 \left( v_u^2 + v'^2_{-} + v'^2 \right)
+\mu^2 \left( v_u^2 + v_d^2 +v^2_{-}  \right)
-2 \mu \epsilon \left( v' v_d + v_{-} v'_{-} \right)\nn\\[+1mm]
&&
+\frac{g^2}{8} \left[
\left( v_u^2-v_d^2-v'^2 +v_{-}^2 +v'^2_{-} \right)^2
+ 4 \left(v_d v_{-} + v' v'_{-} \right)^2 \right] \nn\\[+1mm]
&&
+\frac{g'^2}{8} \left( v_u^2-v_d^2-v'^2 -v_{-}^2 -v'^2_{-} \right)^2
\end{eqnarray}
where $ B'$ characterizes the soft supersymmetry and R--parity
violating bilinear term.  We note that it is not possible to have an
UFB direction with non vanishing charged vev's in this potential,
because the D--terms can not be made to vanish for $v_{-}$ and
$v'_{-}$ different from zero.  The minimization equations can be found
in the usual way taking derivatives with respects to the fields
\begin{eqnarray}
  \label{eq:18}
  0&=&
\left[ 2 \left(M^2_{H_d} + \mu^2\right) -\frac{g^2}{2} \left( 
v_u^2 -v_d^2 -v'^2 -v_{-}^2 +v'^2_{-}  \right) -\frac{g'^2}{2} \left( 
v_u^2 -v_d^2 -v'^2 -v_{-}^2 -v'^2_{-}  \right) 
\right] v_d \nn\\
&&
-\left( 2 \epsilon \mu -g^2 v_{-} v'_{-}\right) v'
-2 B \mu v_u\nn\\[+2mm]
  0&\hskip-3mm=&\hskip-3mm 
\left[ \frac{g^2}{2} \left( 
v_u^2 -v_d^2 -v'^2 +v_{-}^2 +v'^2_{-}  \right)
+\frac{g'^2}{2} \left( 
v_u^2 -v_d^2 +v'^2 -v_{-}^2 -v'^2_{-}  \right) 
\right] v_u \nn\\
&&
+ 2 \left(M^2_{H_u} + \mu^2 + \epsilon^2 \right) v_u + 
2 \left( B' \epsilon v' -B \mu v_d \right)\nn\\[+2mm]
  0&\hskip-3mm=&\hskip-3mm 
\left[ 2 \left(M^2_{L} + \epsilon^2\right) -\frac{g^2}{2} \left( 
v_u^2 -v_d^2 -v'^2 +v_{-}^2 -v'^2_{-}  \right) -\frac{g'^2}{2} \left( 
v_u^2 -v_d^2 -v'^2 -v_{-}^2 -v'^2_{-}  \right) 
\right] v' \nn\\
&&
-\left( 2 \epsilon \mu -g^2 v_{-} v'_{-}\right) v_d
+2 B' \epsilon v_u\nn\\[+2mm]
  0&\hskip-3mm=&\hskip-3mm 
\left[ 2 \left(M^2_{H_d} + \mu^2\right) +\frac{g^2}{2} \left( 
v_u^2 +v_d^2 -v'^2 +v_{-}^2 +v'^2_{-}  \right) -\frac{g'^2}{2} \left( 
v_u^2 -v_d^2 -v'^2 -v_{-}^2 -v'^2_{-}  \right) 
\right] v_{-} \nn\\
&&
-\left( 2 \epsilon \mu -g^2 v_d v' \right) v'_{-} \nn\\[+2mm]
  0&\hskip-3mm=&\hskip-3mm 
\left[ 2 \left(M^2_{L} + \epsilon^2\right) +\frac{g^2}{2} \left( 
v_u^2 -v_d^2 +v'^2 +v_{-}^2 +v'^2_{-}  \right) -\frac{g'^2}{2} \left( 
v_u^2 -v_d^2 -v'^2 -v_{-}^2 -v'^2_{-}  \right) 
\right] v'_{-} \nn\\
&&
-\left( 2 \epsilon \mu -g^2 v_d v' \right) v_{-} 
\end{eqnarray}
Since we are dealing with a set of five coupled equations this system
is difficult to solve for the vev's. We can however use the following
trick.  Instead of solving for the five vev's we try to solve those
equations for the three soft masses squared $M^2_{H_u}$, $M^2_{H_d}$
and $M^2_{L}$~\cite{Romao:1992vu} and for the charged vev's. Using
this approach we could find two types of solutions.

Before discussing the general case, however, we consider first the
limit in which RMSSM is considered a perturbation of the MSSM.  This
is a reasonable approach since the BRpV parameters must be small to
account for the neutrino data~\cite{hirsch:2000ef}. Therefore we can
pose the following question. Suppose that in the limit $\epsilon \to
0$ the parameters are such that the MSSM has no UFB directions or CCB
minima. This means $v_u\not=0\, , v_d\not=0$ and $v'=v_{-}=v'_{-}=0$.
If we now consider a small non--vanishing value for the $\epsilon$
what will be the corresponding minimum? In order to answer this
question in perturbation theory we write
\begin{equation}
  \label{eq:23}
  v_d=\sum_{i=0}^{\infty} v_d^{(i)}\, \epsilon^i, \
  v_u=\sum_{i=0}^{\infty} v_u^{(i)}\, \epsilon^i, \
  v'=\sum_{i=0}^{\infty} v'^{(i)}\, \epsilon^i, \
  v_{-}=\sum_{i=0}^{\infty} v_{-}^{(i)}\, \epsilon^i, \
  v'_{-}=\sum_{i=0}^{\infty} v{'}_{-}^{(i)}\, \epsilon^i 
\end{equation}
Now we substitute back in the extremum Eq.~(\ref{eq:18}) and solve
order by order in perturbation theory. The result that we get is as
follows, 
\begin{eqnarray}
  \label{eq:24}
  v_d&=&v_d^{(0)} +v_d^{(2)}\epsilon^2 +v_d^{(4)}\epsilon^4 + \cdots\nn\\
  v_u&=&v_u^{(0)} +v_u^{(2)}\epsilon^2 +v_u^{(4)}\epsilon^4 + \cdots\nn\\
  v'&=&v'^{(1)} \epsilon +v'^{(3)}\epsilon^3 +v'^{(5)}\epsilon^5 + \cdots\nn\\
  v_{-}&=& 0\nn\\
  v'_{-}&=& 0
\end{eqnarray}
where $v_u^{(0)}\, ,v_d^{(0)}$ are the MSSM values for $\epsilon=0$.
This is precisely the solution of type \textbf{I} that we will discuss
shortly.  Note that if $\epsilon \not=0$ then also $v'\not=0$. In
fact,
\begin{equation}
  \label{eq:25}
  v'= \frac{\mu v_d^{(0)} 
- B' v_u^{(0)}}{M^2_L- \frac{1}{4} (g^2 + g'^2)
\left(v_u^{(0)}{}^2 - v_d^{(0)}{}^2\right)}\, \epsilon +\cdots
\end{equation}
So we can formulate the following important result: {\it If we start
  with the MSSM parameters such that in the limit $\epsilon
  \rightarrow 0$ the minimum has no UFB or CCB problems, then by
  turning on perturbatively a small value for $\epsilon$ we get a
  correspondingly safe minimum of the RMSSM.} However, as we will now
discuss, in general there are two types of solutions for the minimum
equations.

\subsubsection*{Type I}
This solution corresponds to the case where the charged vev's vanish.
We are then in the situation studied usually \cite{Diaz:1997xc} in the
bilinear R--parity model. We get

\begin{eqnarray}
  \label{eq:19}
  M^2_{H_d}&=& 
\epsilon\, \mu\, \frac{v'}{v_d} - \mu^2 + B\, \mu\, \frac{v_u}{v_d} 
+ \frac{g^2 + g'^2}{4} \left(v_u^2 -v'^2 - v_d^2 \right)
  \nn\\[+2mm]
  M^2_{H_u}&=& 
-\epsilon^2\, - \mu^2 + B\, \mu\, \frac{v_d}{v_u}
 -B'\, \epsilon\, \frac{v'}{v_u} 
- \frac{g^2 + g'^2}{4} \left(v_u^2 -v'^2 - v_d^2 \right)
  \nn\\[+2mm]
  M^2_{L}&=& 
-\epsilon^2 +\epsilon \left(\mu \frac{v_d}{v'}- B' \frac{v_u}{v'}\right) 
+ \frac{g^2 + g'^2}{4} \left(v_u^2 -v'^2 - v_d^2 \right)
  \nn\\[+2mm]
  v_{-}&=&0
  \nn\\[+2mm]
  v'_{-}&=&0
\end{eqnarray}
This corresponds to the neutral Higgs potential that we will discuss
further below. Here we just note that the value of the potential at 
the minimum can be shown to be
\begin{equation}
  \label{eq:26}
  V_{BRpV}= -\frac{g^2+g'^2}{8}
  \left(v_u^2-v_d^2-v'^2\right)^2.
\end{equation}

\subsubsection*{Type II}

In the general case we can find the solutions of the minimization
equations in the following way. We start by solving the first three
equations in Eq.~(\ref{eq:18}) for the soft masses. We get,
\begin{eqnarray}
  \label{eq:20}
  M^2_{H_d}&=& M^2_{H_d}(0) -\frac{1}{4} \left(g^2-g'^2\right)
  \left(v_{-}^2+ v'^2_{-}\right) \nn\\
  M^2_{H_u}&=& M^2_{H_u}(0) -\frac{1}{4} \left(g^2+g'^2\right)
  \left(v_{-}^2+ v'^2_{-}\right) 
  -\frac{1}{2}\, g^2\, \frac{v'}{v_d}\, v'_{-} v_{-}\nn\\
  M^2_{L}&=&M^2_{L}(0) +\frac{1}{4} \left(g^2-g'^2\right) v_{-}^2
-\frac{1}{4} \left(g^2+g'^2\right)  v'^2_{-}
  -\frac{1}{2}\, g^2\, \frac{v'}{v_d}\, v'_{-} v_{-}
\end{eqnarray}
where $M^2_{H_d}(0)$, $M^2_{H_u}(0)$ and $M^2_{L}(0)$ are the soft
masses when $v_{-}=v'_{-}=0$ and are given in
Eq.~(\ref{eq:19}). Now we substitute Eq.~(\ref{eq:20}) into the last
two equations in Eq.~(\ref{eq:18}) to obtain,
\begin{eqnarray}
  \label{eq:35}
  0&\hskip-3mm=\hskip-3mm&
-g^2\left( v'^2 v_{-} - v' v_d v'_{-} + 
 v_{-}^2 v'_{-} \frac{v'}{v_d} - v_{-} v'^2_{-} \right)
 + 2 \epsilon \mu \left( v_{-}\frac{v'}{v_d} - v'_{-} \right) 
+ 2 B \mu v_{-} \frac{v_u}{v_d} + g^2 v_{-} v_u^2\nn\\[+2mm]
0&\hskip-3mm=\hskip-3mm&
 g^2 \left( v'^2 v_{-} 
- v' v_d v'_{-} +  v_{-}^2 v'_{-}  \frac{v'}{v_d}
- v_{-}  v'^2_{-} \right) \frac{v_d}{v'}
 - 2 \epsilon \mu \left( v_{-}\frac{v'}{v_d} - v'_{-} \right) 
\frac{v_d}{v'} \\
&&
- 2 B' \epsilon v'_{-} \frac{v_u}{v'} + g^2 v'_{-} v_u^2\nn
\end{eqnarray}
Multiplying the second of the equations in Eq.~(\ref{eq:35}) by
$v'/v_d$ and adding them one obtains,
\begin{equation}
  \label{eq:36}
  v'_{-}=\kappa\, v_{-}
\end{equation}
where
\begin{equation}
  \label{eq:37}
  \kappa=\frac{2 B \mu + g^2 v_d v_u}{2 B' \epsilon - g^2 v' v_u}  
\end{equation}

\vspace{1mm}

\ni
Finally we use Eq.~(\ref{eq:36}) to reduce either one of Eq.~(\ref{eq:35})
to
\begin{equation}
  \label{eq:38}
  0=v_{-} \Big( D_2\, v_{-}^2 - D_0 \Big)
\end{equation}
where
\begin{eqnarray}
  \label{eq:39}
  D_2&=& g^2 \left( \kappa^2 - \frac{v'}{v_d}\, \kappa \right)\nn\\
  D_0&=&g^2 \left( v'^2 -v_d v' \kappa -v_u^2 \right)
- \left( B v_u + \epsilon v'\right) \frac{2 \mu}{v_d}
+ 2 \epsilon \mu \kappa
\end{eqnarray}
Eq.~(\ref{eq:38}) has the trivial solution $v_{-}=0$ which corresponds
to type \textbf{I}, the BRpV solutions. However, if
\begin{equation}
  \label{eq:22}
  \frac{D_0}{D_2} > 0
\end{equation}
we have a new type of solutions for the minimization equations,
\begin{equation}
  \label{eq:40}
  v_{-}=\pm \sqrt{\frac{D_0}{D_2}}, \qquad v'_{-}= \kappa\, v_{-}
\end{equation}
As $D_{0,2}$ do not have in general a well defined sign it can happen
that such solutions do exist for some combination of the
parameters. We will discuss this later in more detail. 

\subsection{UFB Directions}

We have seen before that for the Higgs potential of the RMSSM the UFB
directions can only arise when the charged Higgs vev's vanish,
otherwise it is not possible to cancel the quartic D--terms.  The
neutral Higgs potential obtained from Eq.~(\ref{eq:15}) when
$v_{-}=0,v'_{-}=0$ is given by
\begin{eqnarray}
  \label{eq:27}
  V_{Neutral}&=& \left( M^2_{H_u} +\epsilon^2 +\mu^2 \right) v_u^2 
  + \left(M^2_{H_d} +\mu^2 \right) v_d^2
  + \left(M^2_{L} + \epsilon^2 \right) v'^2  \nn\\[+1mm]
&&- 2 B \mu\, v_d v_u + 2 B' \epsilon\, v_u v' 
-2 \mu \epsilon v' v_d 
+\frac{g^2+g'^2}{8} 
\left( v_u^2-v_d^2-v'^2\right)^2
\end{eqnarray}
From this equation we can see that we can make the D--term  vanish if
we choose the condition
\begin{equation}
  \label{eq:28}
  v_u^2=v_d^2+v'^2
\end{equation}
To implement this condition it is convenient to write
\begin{equation}
  \label{eq:29}
  v_d=v_u \cos\theta, \quad v'=v_u\sin\theta
\end{equation}
Then we get
\begin{equation}
  \label{eq:30}
  V_{Neutral}=B(\theta) v_u^2
\end{equation}
where
\begin{eqnarray}
  \label{eq:31}
  B(\theta)&=& 
\left[\vb{14} M^2_{H_u} +\epsilon^2 +\mu^2 
  + \left(M^2_{H_d} +\mu^2 \right) \cos^2\theta
  + \left(M^2_{L} + \epsilon^2 \right) \sin^2\theta  \right.\nn\\[+1mm]
&&\left.\vb{14} \hskip 3mm
- 2 B \mu\, \cos\theta + 2 B' \epsilon\, \sin\theta 
-2 \mu \epsilon \sin\theta \cos\theta \right]
\end{eqnarray}
Therefore the condition for avoiding an UFB direction is that,
\begin{equation}
  \label{eq:32}
  B(\theta_{\min}) > 0
\end{equation}
where $\theta_{\min}$ is the value of $\theta$ that corresponds to the
minimum of $B(\theta)$.  Now consider Eq. (\ref{eq:31}) in the limit
$\epsilon \to 0$ and take the derivative,
\begin{equation}
\label{eq:Btheta}
\frac{dB}{d\theta}= 2\sin\theta \left[-( M^2_{H_d} +\mu^2-M^2_{L})\cos\theta +   
B \mu\ \right]
\end{equation}
The right hand side vanishes when $\theta=0$ and when $\cos\theta =
\frac{B\mu}{M^2_{H_d} +\mu^2 - M^2_{L}}$. These two solutions
correspond to the UFB-1 and UFB-2 directions given in
Eqs.~(\ref{eq:11}) and (\ref{eq:ufb2}), respectively, when
$\epsilon=0$.

\begin{figure}[ht]
  \centering
  \begin{tabular}{cc}
    \includegraphics[width=0.45\textwidth,clip]{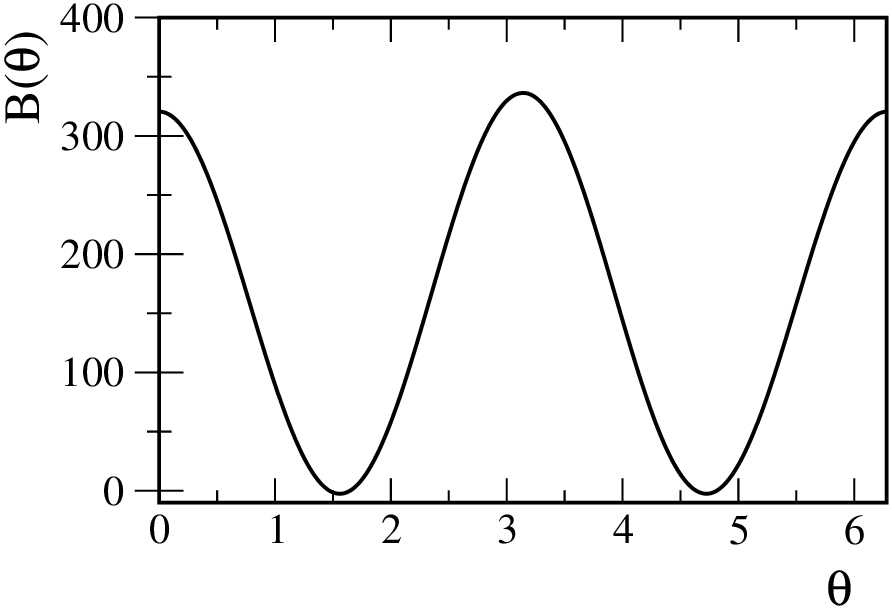}&
    \includegraphics[width=0.45\textwidth,clip]{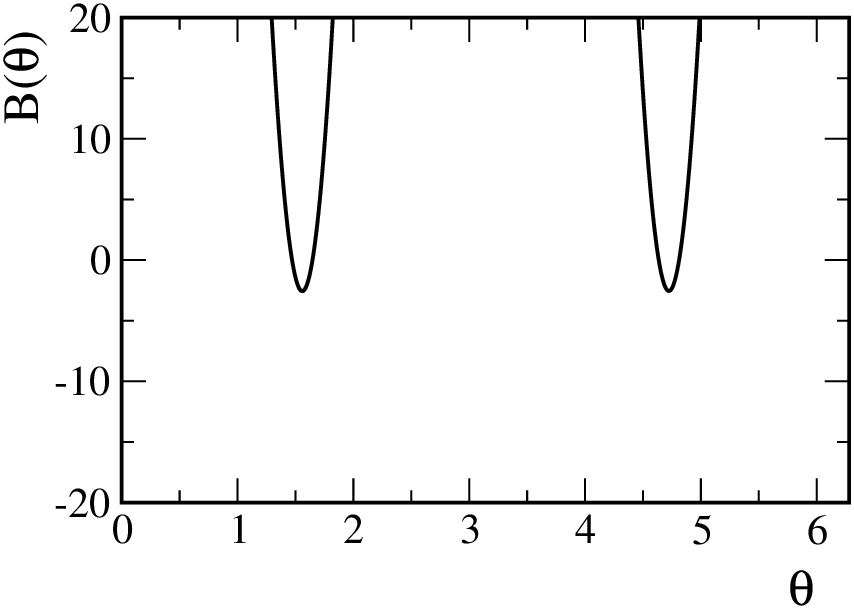}
  \end{tabular}
  \caption{$B(\theta)$ as a function of $\theta$ for an example where
    $B(\theta_{min})<0$ but $B(0)>0$. The right panel is an enlarged
    view of the left one close to the zeros of  $B(\theta)$.}
  \label{fig:1}
\end{figure}

For $\epsilon \not=0$ it does not seem possible to have an analytical
expression for $\theta_{\min}$. However for a given set of parameters
it is always easy to verify whether Eq.~(\ref{eq:32}) holds for
$\theta \in [0, 2 \pi]$. It is also clear from Eq.~(\ref{eq:31}) that
the MSSM condition, Eq.~(\ref{eq:11}), is not enough to ensure that we
are free from UFB directions. This fact can be best illustrated from
figure (\ref{fig:1}) that shows a typical example.

One can see clearly that starting from a large value of $B(0)$ is not
enough to decide upon the sign of $B(\theta_{\min})$. However it is
easy to check numerically whether $B(\theta_{\min})>0$ or not.
Therefore, although we lack a simple analytical formula, the criterium
for avoiding UFB directions is easily implemented.

Finally we comment briefly on the direction UFB-3. It can be easily
shown that at large values of the field the potential in direction
UFB-3 is given as
\begin{equation}
  \label{eq:ufb3mod}
  V_{UFB-3}=\left( m^2_{H_u} + m^2_{L_i} + \epsilon B' \right) v_u^2 
  + \cdots
\end{equation}
where the dots stand for irrelevant terms. Since in our notation
$\epsilon B' < 0$ this leads, in principle, to a slightly more
stringent requirement than the one corresponding to the R--parity
conserving MSSM. However, since $\frac{\epsilon}{\mu} \sim {\cal
  O}(10^{-(3-4)})$ is required by neutrino oscillation data
\cite{hirsch:2000ef}, this modification is numerically irrelevant.
This is in agreement with the argument presented in
Ref.~\cite{Abel:1998ie}.

\subsection{Nonzero charged Higgs and Slepton Vev's}

We now turn to the solutions of type II. We have already seen in Eqs.
(\ref{eq:38}) - (\ref{eq:40}) that there are potentially dangerous
solutions for the Higgs potential with nonzero vev's for the charged
scalars. These solutions, if they exist, would provide new CCB
solutions different from those already present in the MSSM, as
explained above.  As can be seen from Eq.~(\ref{eq:22}) such solutions
can exist if the parameters satisfy the relation $D_0/D_2 >0$, where
the $D_i$ are given in Eq.~(\ref{eq:39}).

Since it does not seem possible to give a strict analytic criterion
which relates the condition $D_0/D_2 <0$ (guaranteeing the absence of
unwanted minima) to the parameters of the potential we have resorted
to a numerical scan of the parameter space.  Our approach to find the
minima of the potential was as follows. We always started with a
random set of parameters with zero charged vev's and subject to the
requirement that,
\begin{equation}
  \label{eq:41}
  v_u^2+v_d^2+v'^2=v^2=\left( 2 \sqrt{2} G_F\right)^{-1/2}= 174.1\,
  \mathrm{GeV} 
\end{equation}

\begin{figure}[ht]
  \centering
  \begin{tabular}{cc}
    \includegraphics[width=0.45\textwidth,clip]{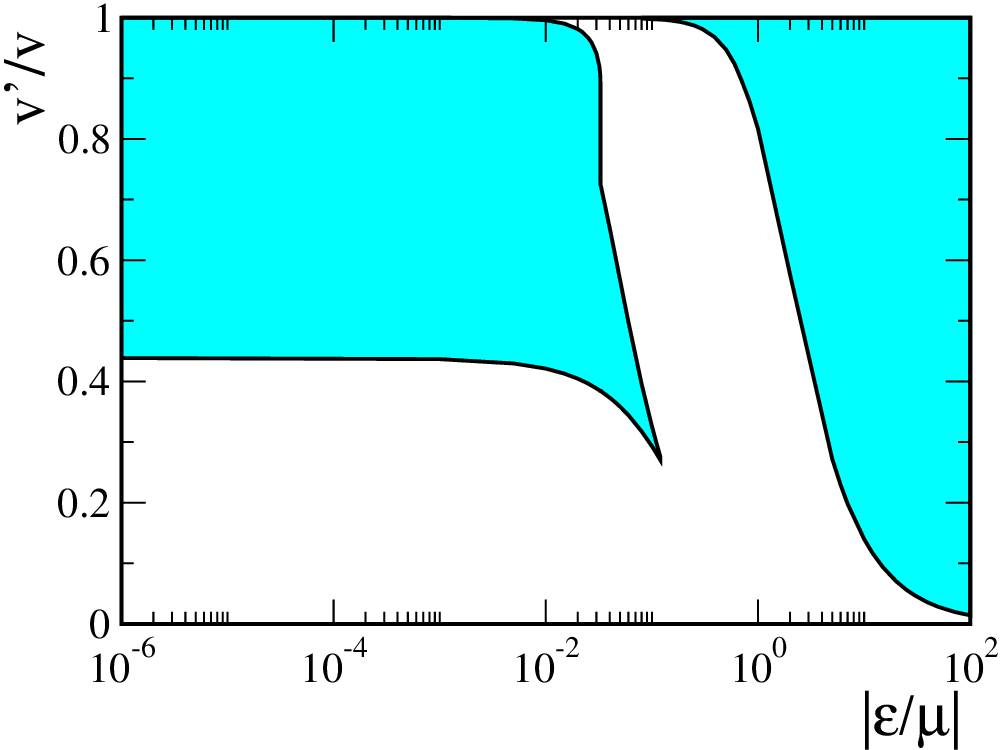}&
    \includegraphics[width=0.45\textwidth,clip]{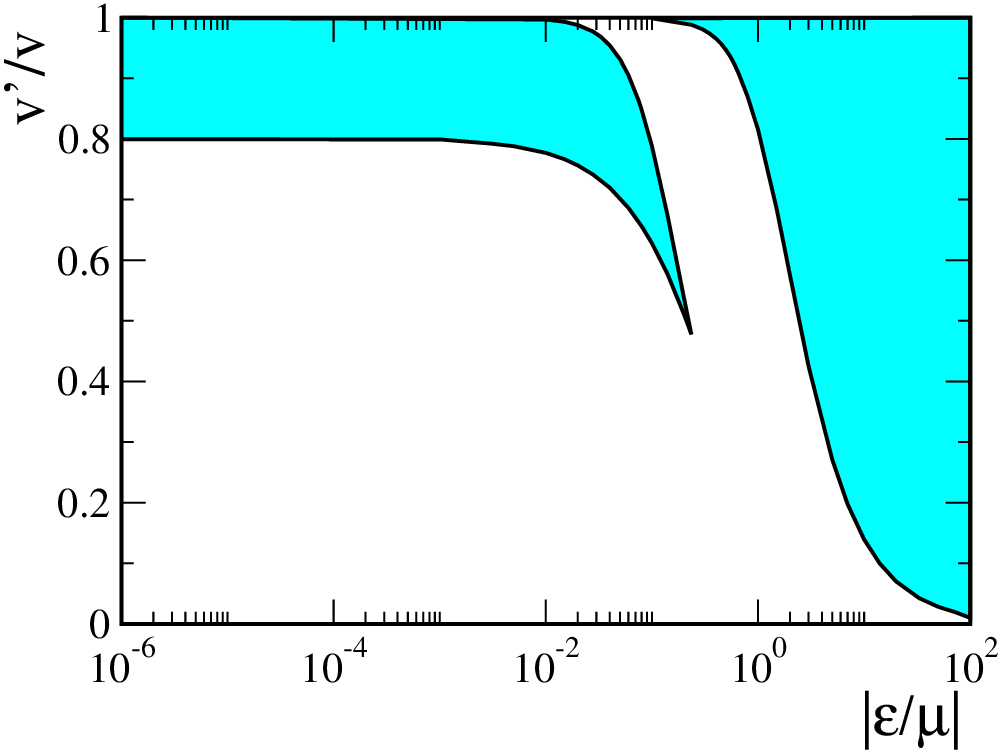}
  \end{tabular}
  \caption{Range of RMSSM parameters where 
    nonzero charged vev's for the Higgs and slepton fields are
    favoured over the realistic minimum for two examples of
    $\tan\beta$, left $\tan\beta=1.05$, right $\tan\beta=1.2$. Here we
    fix for convenience $B=B'=\mu=$ 100 GeV. For a discussion see
    text.}
  \label{fig:2}
\end{figure}

\ni Note that with this procedure we should always have,
\begin{equation}
  \label{eq:42}
  |\eta| =\frac{|v'|}{v} < 1.
\end{equation}
We then search for the global minimum numerically. If we find a minimum 
deeper than the realistic minimum but which breaks charge this part of 
parameter space should be discarded.  
Two examples are shown in Fig. (\ref{fig:2}). 

The results shown in Fig. (\ref{fig:2}) can be understood
qualitatively as follows. Starting with the definitions Eqs
(\ref{eq:37}) and (\ref{eq:39}) and taking into account the smallness
of $\frac{\epsilon}{\mu}$ one can show that in the limit $\epsilon \to
0$ we always have $D_2 >0$. On the other hand the condition $D_0>0$
requires
\begin{equation}
\label{eq:condd}
v'^2 > v^2\frac{\tan^2\beta-1}{1+\tan^2\beta} + \frac{2 B \mu}{g} 
\frac{\tan^2\beta-1}{\tan\beta}.
\end{equation}
Note that this condition is not strictly valid for $\tan\beta \equiv
1$, because in this limit we can no longer neglect the terms
proportional to $\epsilon$ in the definitions of $D_0$ and $D_2$.  Eq.
(\ref{eq:condd}) shows that charge breaking minima in the limit of
small values of $\epsilon$ require that $v'$ take up a sizeable
fraction of $v$. This trend is clearly visible from Fig.
(\ref{fig:2}).  The figure also illustrates how these solutions
disappear very quickly with $\tan\beta$ greater than $1$.

Although we find it amusing that such solutions exist, we wish to
stress that consistency with neutrino data requires
$\frac{\epsilon}{\mu} \sim {\cal O}(10^{-(3-4)})$ and $\frac{v'}{v}
\sim {\cal O}(10^{-(3-4)})$. We therefore conclude that the RMSSM is
automatically safe from these unwanted minima in those ``physical''
parts of parameter space which account for the neutrino oscillation
data.

\section{Conclusions}

We have studied charge breaking minima and unbounded from below
directions within bilinear R--parity breaking supersymmetry. Such a 
``reference model'' is nothing but the simplest broken R--parity
version of the Minimal Supersymmetric Standard Model.  We have first
generalized some results obtained previously in the R--parity
conserving MSSM. Subsequently we discussed new ways to generate a
nonzero vacuum expectation value of the charged Higgs and slepton
fields.  However, such unwanted solutions occur only in regions of
parameter space which are now excluded by neutrino oscillation data.

In summary it can be said that, given the data on neutrino masses,
bilinear R--parity violation can be understood as a small perturbation
of the MSSM. From the point of view of CCB and UFB directions the
RMSSM is as robust as the R--parity--conserving MSSM: it is equally
safe from unwanted minima in the same portions of parameter space.

\section{Acknowledgments}

This work was supported by Spanish grant BFM2002-00345, by the
European Commission Human Potential Program RTN network
HPRN-CT-2000-00148 and by the European Science Foundation network
grant N.86.  M.H. is supported by a MCyT Ramon y Cajal contract.
JCR was supported by the Portuguese 
\textit{Funda\c{c}\~ao para a Ci\^encia e a Tecnologia} 
under the contract CFIF-Plurianual and grant POCTI/FNU/4989/2002. 
We thank Werner Porod for useful discussions.

\end{document}